\documentclass[pre,aps,twocolumn]{revtex4}

\usepackage{graphicx}
\usepackage{amsmath,empheq}
\usepackage{amssymb}
\usepackage{ifthen}
\usepackage{booktabs}

\usepackage{graphicx}
\usepackage{dcolumn}
\usepackage{bm}
\usepackage{longtable}
\usepackage{gensymb}

\newcommand{\bb}{\begin{equation}}
\newcommand{\ee}{\end{equation}}
\newcommand{\ba}{\begin{eqnarray*}}
\newcommand{\ea}{\end{eqnarray*}}
\newcommand{\rhor}{\rho({\bf r})}

\newcommand{\rr}{{\mathbf r}}

\begin{document}

\title{Three-phase fluid coexistence in heterogenous slits}

\author{Martin \surname{L\'aska}}
\affiliation{
{Department of Physical Chemistry, University of Chemical Technology Prague, Praha 6, 166 28, Czech Republic;}\\
 {Department of Molecular and Mesoscopic Modelling, ICPF of the Czech Academy Sciences, Prague, Czech Republic}}                
\author{Andrew O. \surname{Parry}}
\affiliation{Department of Mathematics, Imperial College London, London SW7 2BZ, UK}
\author{Alexandr \surname{Malijevsk\'y}}
\affiliation{ {Department of Physical Chemistry, University of Chemical Technology Prague, Praha 6, 166 28, Czech Republic;}
 {Department of Molecular and Mesoscopic Modelling, ICPF of the Czech Academy Sciences, Prague, Czech Republic}}

\begin{abstract}
\noindent  We study the competition between local (bridging) and global condensation of fluid in a chemically heterogeneous capillary slit made from
two parallel adjacent walls each patterned with a single stripe. Using a mesoscopic modified Kelvin equation, which determines the shape of the
menisci pinned at the stripe edges in the bridge phase, we determine the conditions under which the local bridging transition precedes capillary
condensation as the pressure (or chemical potential) is increased. Provided the contact angle of the stripe is less than that of the outer wall we
show that triple points, where evaporated, locally condensed and globally condensed states all coexist are possible depending on the value of the
aspect ratio $a=L/H$ where $H$ is the stripe width and $L$ the wall separation. In particular, for a capillary made from completely dry walls
patterned with completely wet stripes the condition for the triple point occurs when the aspect ratio takes its maximum possible value $8/\pi$. These
predictions are tested using a fully microscopic classical Density Functional Theory and shown to be remarkably accurate even for molecularly narrow
slits. The qualitative differences with local and global condensation in heterogeneous cylindrical pores are also highlighted.
\end{abstract}

\maketitle

\begin{figure}
\includegraphics[width=7cm]{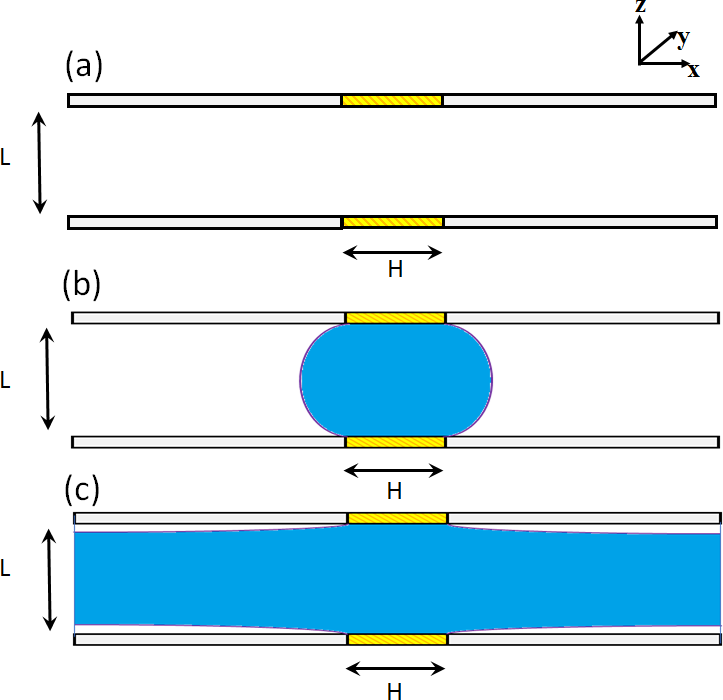}
\caption{Illustration of a) gas-like, b)  locally condensed bridge-like  and c) liquid-like  states in a chemically heterogenous slit. Possible
configurations depend on $\mu$.}
\end{figure}

Confined  fluids can exhibit dramatically different phase behaviour and associated phase transitions to that occurring in the bulk \cite{dietrich,
schick, hend}. Even for a single wall (substrate), the presence of the wall-fluid interaction induces rich new surface phase phenomena: wetting
transitions are a very well known example of this \cite{cahn, ebner, ev_tar, lip}. The complexity of the fluid phase behaviour is further enriched if
the wall is geometrically or chemically heterogeneous \cite{bonn}. Wedge filling \cite{hauge, rejmer, wood1, our_wedge}, step wetting \cite{saam},
bridging \cite{lenz, bauer, posp} and the change from Cassie's \cite{cassie} to Wenzel's \cite{wenzel} state are but a few examples of phase
transitions directly induced by the substrate geometry. Another related phenomenon is capillary condensation referring to the shift of the bulk
liquid-gas phase boundary, when a fluid is confined between two parallel walls separated by a distance $L$. Macroscopically, the shift in the
chemical potential, relative to saturation, at which capillary condensation (cc) occurs is given by Kelvin's equation
  \bb
  \delta \mu_{cc}\equiv\mu_{\rm sat}-\mu_{cc}=\frac{2\gamma\cos\theta}{L\Delta\rho}\,,\label{cc}
 \ee
where $\gamma$ is the liquid-gas surface tension, $\theta$ is Young's contact angle and $\Delta\rho$ is the difference between number densities of
coexisting liquid and gas. Here, the Laplace pressure $\delta p$ due the curved interface separating liquid and gas phases has been approximated by
$\delta p\approx\delta\mu_{cc}\Delta\rho$ valid for small undersaturation \cite{evans90}. Studies based on microscopic Density Functional Theory
(DFT) models \cite{ebner, evans79} have shown that the Kelvin equation is remarkably robust. This is particularly true for partial wetting
($\theta>0$), where thick wetting layers are absent at each wall.

There have also been a number of studies of fluid confinement between walls that are themselves geometrically sculpted or patterned, that is when
translational invariance is broken both across and along the slit \cite{chmiel, rocken, frink, bock, swain, bock2, valencia, hemming, schoen}. In
particular numerical DFT studies have established that capillary condensation, which must still be present, competes with other types of phase
transition induced by the patterning or equivalently that the condensation occurs in two steps. In this paper we show that this problem can be
studied analytically and derive results characterizing the two step condensation which apply to capillaries of all scales from the macroscopic down
to the molecular. To do this we consider the case in which each wall is chemically heterogeneous and has a macroscopically long stripe or patch of
width $H$ for which the local contact angle is $\theta_p$. Outside of this region the wall is made from a different material where the contact angle
is $\theta$. The stripes are adjacent and parallel (see Fig.~1). Since the outer wall area is macroscopically large the location of the global
capillary condensation from a gas-like to a liquid-like phase as the chemical potential is increased is unaffected by the presence of the stripes.
However, if $\theta_p<\pi/2$ (so there is greater affinity to liquid than to gas) {\it and} $\theta_p<\theta$, so that it preferentially adsorbs the
liquid relative to the outer wall, it is also possible that the fluid locally condenses forming a liquid bridge between the patches. This is a third
possible phase for the confined fluid which competes with the evaporated (gas-like) and globally condensed (liquid-like) fluid states (see Fig.~1).
Here we establish the conditions under which local condensation (bridging) precedes capillary condensation as the chemical potential is increased
towards saturation. Our analysis uses a modified Kelvin equation not considered in previous works which allows for an edge contact angle $\theta_E$
which describes the shape of the menisci pinned at the edges of both patches. Our main new result is that for a slit maximum wetting contrast in
which a completely wet stripe is embedded within completely dry wall, the value of the aspect ratio determining the triple point is maximum and takes
a universal value $8/\pi$. This prediction is tested using microscopic DFT and is shown to be extraordinarily accurate even for molecularly narrow
slits.

Let us begin with mesoscopic considerations. As mentioned above since the outer wall area is infinitely greater than that of the patch the global
condensation from a gas-like to liquid-like state occurs at the same chemical potential $\delta\mu_{cc}$ identified by the standard Kelvin equation.
We next need to determine the chemical potential at which a gas-like phase locally condenses forming a bridge between the patches. Here we use a
modified Kelvin equation similar to one recently proposed for opened slits. This, bridge phase, is characterized by two circular menisci of Laplace
radius $R=\gamma/\delta p$ which separate the liquid within the bridge from the gas in the outer regions of the capillary. These menisci are
necessarily pinned at the edges of the patch  and meet the walls with an edge contact angle $\theta_E$ whose value depends on the thermodynamic
state, as well as the properties of the heterogenous slit. However, exactly at the bridging transition, the value of $\theta_E$ depends only on the
contact angle of the patch $\theta_p$ and the aspect ratio $a\equiv L/H$; there is no dependence on the outer contact angle $\theta$ since the
exposed wall surface area to the vapour phase is the same for the evaporated and bridge phases.

The situation is closely analogous to condensation in an open slit and balancing the bulk and surface free energy contributions  of the evaporated
and bridge phases determines that a local condensation to the bridge phase occurs at the chemical potential $\mu_{b}\equiv \mu_{\rm sat}(T)-\delta
\mu_{b}$ where the shift is determined by the modified Kelvin equation
 \bb
  \delta \mu_{b}=\frac{2\gamma\cos\theta_E}{L\Delta\rho}\,,\label{bridge}
 \ee
Here $\theta_E$ is the edge contact angle whose value at the bridging transition is given by \cite{openslit, supmat}
 \bb
 \cos\theta_p=\cos\theta_E+\frac{a}{2}\left[\sin\theta_E+\sec\theta_E\left(\frac{\pi}{2}-\theta_E\right)\right]\,. \label{thetae}
 \ee
which, as mentioned above, is independent of the properties of the outer wall.

Comparing the chemical potential shifts (\ref{cc}) and (\ref{bridge}) immediately determines what phases are stable and metastable in the
heterogeneous capillary slit. It follows that on increasing the chemical potential a bridging transition precedes the global condensation only if
$\theta_E<\theta$ since the chemical potential shift $\delta \mu_b>\delta \mu_{cc}$. In this case on increasing $\mu$ the gas-like phase in capillary
first locally condenses (bridges the walls) near the patch at the chemical potential $\mu_b$. On further increasing the chemical potential the bridge
phase eventually condenses to a liquid-like phase when the chemical potential reaches $\mu_{cc}$. During this process the value of the edge contact
angle increases continuously from its value given by (\ref{thetae}) at $\mu_b$ taking the value $\theta$ exactly at $\mu_{cc}$. This means that at
the condensation the menisci are no longer pinned at the edges by the macroscopic forces of tension i.e. translating the menisci away from the patch
does not alter the macroscopic free energy. We return to this later in our discussion of fluctuation effects. On the other hand if $\theta_E>\theta$
the global condensation occurs prior to the bridging which is only ever a possible metastable phase. These two possibilities are separated by the
case $\theta_E=\theta$ which is the condition for a triple point where all three phases coexist.

It follows that for given contact angles $\theta_p$ and $\theta$, the value of the aspect ratio $a_T$ at which a triple point occurs is given by
 \bb
  a_T=\frac{2(\cos\theta_p-\cos\theta)}{\sin\theta+\left(\frac{\pi}{2}-\theta\right)\sec\theta}\,. \label{triple}
 \ee
Only if the aspect ratio $a<a_T$ does a bridging transition precede global capillary condensation. Therefore, as anticipated, bridging  is only
possible if the contact angle in the patch $\theta_p<\theta$. It is only in this case that the free energy cost of creating two menisci is
compensated by the greater affinity of the patches for the liquid. We can also identify that the condition for the maximum value of the triple point
aspect ratio. This corresponds to a heterogeneous slit of maximum contrast (MC) comprising a completely wet patch ($\theta_p=0$) embedded within a
completely dry outer wall ($\theta=\pi$). These values of the contact angles minimize the denominator and maximize the numerator in (\ref{triple})
leading to the intriguing value
  \bb
  a_T^{\rm max}=\frac{8}{\pi}\,.\label{triplemc}
   \ee
Thus only for $a<\frac{8}{\pi}$ will a MC slit exhibit  a bridging transition which precedes global capillary condensation. We also note that for
$a<1$ the value of $\mu_b$ at which this occurs is less than $\mu_{\rm sat}$ meaning that the menisci are concave. For aspect ratio $a=1$ the
bridging transition occurs exactly at $\mu_{\rm sat}$ and the menisci are flat. For $a>1$ the menisci are convex and the bridging transition occurs
at $\mu_b>\mu_{\rm sat}$.

We  test these predictions using a microscopic DFT to determine equilibrium density profiles $\rho_{\rm eq}(\rr)$ and the corresponding free energies
of possible stable or metastable states. These are obtained by minimization of the grand potential functional
\begin{equation}
\Omega[\rho]=F[\rho]-\int d{\bf{r}}(\mu-V({\bf{r}}))\rho({\bf{r}})
\end{equation}
where $V({\bf{r}})$ is the external potential exerted by the heterogenous slit and $F[\rho]$ is the intrinsic free-energy functional of the one-body
density $\rhor$. We concentrate on the maximum contrast slit geometry and the prediction (\ref{triple}) for the value of $a_T^{\rm max}$. Explicit
expressions for the corresponding external potential as well as the approximative free-energy functional adopted are shown in the Supplemental
Material \cite{supmat}.

 \begin{figure}
\centerline{\includegraphics[width=7cm]{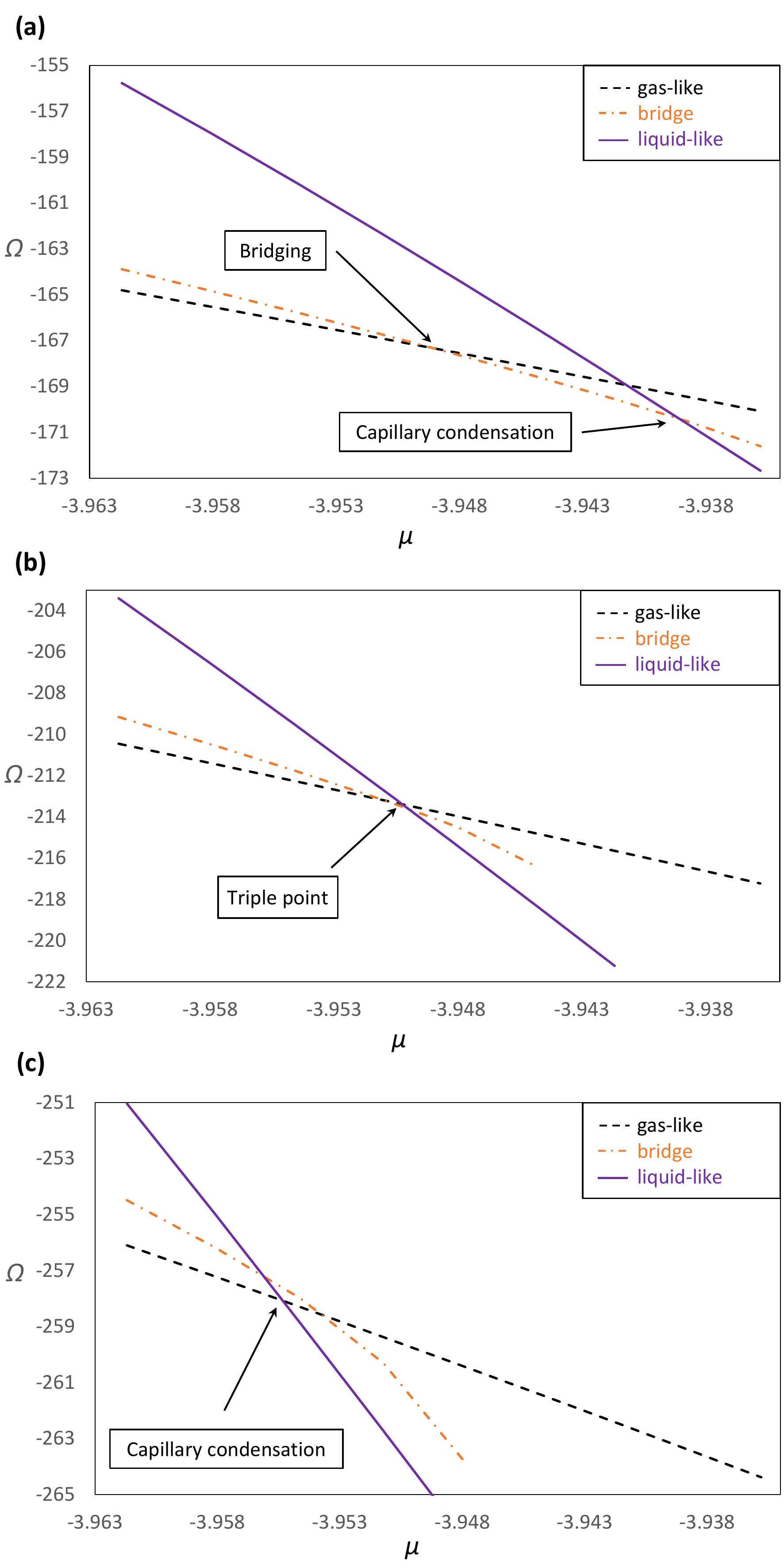}}
 \caption{DFT results for the grand potential as a function of the chemical potential (both in units of $\varepsilon$) for the maximum contrast
 slit with $\theta_p=0$ and $\theta=\pi$ and $H=10\,\sigma$.
Three values of $L$ are shown corresponding to the aspect ratios a) $a=2.3$ for which bridging precedes the global capillary condensation, b) $a=2.6$
our best estimate of the triple point and c) $a=2.8$ where only single capillary condensation transition is stable and the bridge configurations are
always metastable. Our estimate to the triple point $a=2.6$ is very close to the predicted value $a_T$ as given by (\ref{triplemc}). }
\end{figure}

 \begin{figure}
\centerline{\includegraphics[width=7cm]{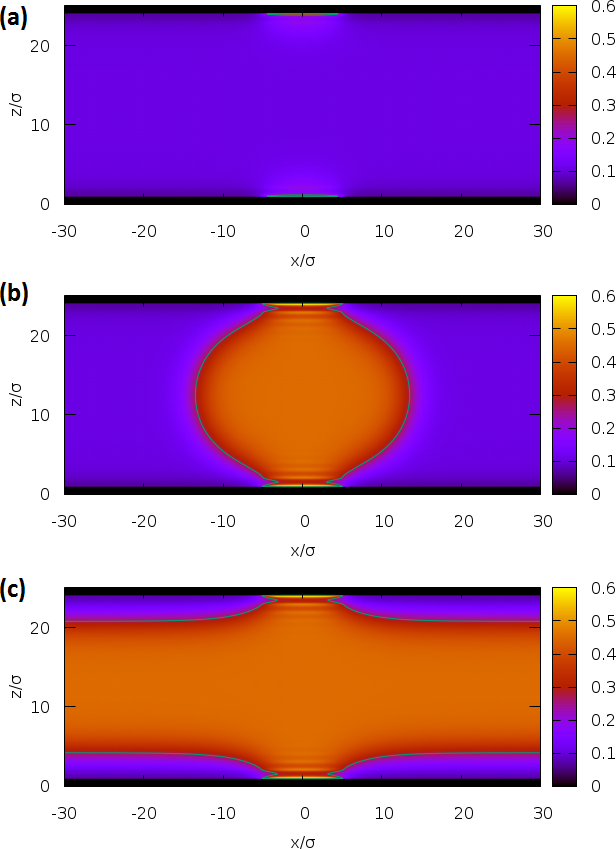}} \caption{Coexisting density profiles for the slit corresponding to the triple point of Fig.~2b showing
a) gas-like, b) bridge and c) liquid-like states.}
\end{figure}

We have considered two different values of the patch width $H=10\,\sigma$ and $H=20\,\sigma$ and for each varied the slit width $L$, thus varying the
aspect ratio  $a$. For a given aspect ratio we  varied the chemical potential and determined the equilibrium stable and metastable phases
corresponding to evaporated, bridged and condensed states. This is achieved by determining numerically the minimal grand potential using Picard's
iteration starting from an initial configuration corresponding to these phases. In Fig.~2 we present results for the grand potential as a function of
the chemical potential for $a=2.3$, $a=2.6$, and $a=2.8$ for the smallest system size $H=10\,\sigma$. For the smallest value of $a$ we observe two
kinks in the minimum value of the grand potential corresponding to bridging and global capillary condensation transitions as the chemical potential
is increased. On increasing the value of $a$ the kinks move closer and eventually merge at a value $a=2.6$ which, even for this microscopically small
patch, is very close to the prediction (\ref{triple}). Density profiles corresponding to the the coexisting phases are shown in Fig.~3. For the
bridge phase the microscopic menisci are indeed convex as anticipated. For the largest value of $a$ shown there is only one kink which corresponds to
single capillary condensation. Bridge states also exist but have a higher grand potential and are therefore metastable. Similar results have been
obtained for larger values of $H$ with the aspect ratio corresponding to the triple point always very close to the predicted value $8/\pi$. We find
it remarkable that the prediction for the triple point value of the aspect ratio based on the Kelvin and modified Kelvin equations is highly accurate
even at the nanoscale.

To finish our paper we mention three further points. Firstly, so far we have supposed that the contact contact angle of the patch is less than that
of the outer wall. If we reverse this scenario, so that $\theta_p>\theta$ and $\theta_p>\pi/2$, then the patch has greater affinity for the gas
compared to the outer wall. In this case, analogous bridging transitions take place but involve a bubble of gas that with two menisci separating the
bubble from the outer liquid. Thereafter the analysis of the bridging and global condensation transitions remain the same including the location of
the triple point, although the numerator in (\ref{triple}) is replaced by $\cos\theta-\cos\theta_E$.

 \begin{figure}
\includegraphics[width=6cm]{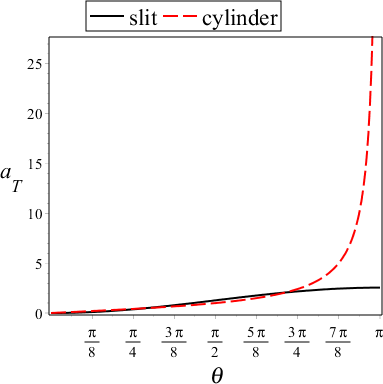}
\caption{Predicted triple-point aspect ratios for the heterogenous slit and the pore geometries for the completely wet patch $\theta_p$ as a function
of the contact angle $\theta$ of the outer wall. }
\end{figure}

Secondly, we mention that qualitatively similar phenomena occur in a heterogenous cylindrical pore of radius $\mathcal{R}$ containing a patch of
width $H$ for which the local contact angle is $\theta_p$. For this we define an aspect ratio $a^{\rm cyl}=2{\mathcal{R}}/H$. Again, assuming that
$\theta_p$ is less than the contact angle $\theta$ of the outer wall, we must consider the possibility of a local bridging transition occurring at
$\delta p=2\gamma\cos\theta_E^{\rm cyl}/{\mathcal{R}}$ and a global condensation occurring at $\delta p=2\gamma\cos\theta/{\mathcal{R}}$. The value
of $\theta_E^{\rm cyl}$ at the transition is again known from previous studies of condensation in open pores \cite{openslit}. The analysis is
analogous to that for the heterogenous slit and setting $\theta_E^{\rm cyl}=\theta$ leads to the condition for the value of the aspect ration
$a_T^{\rm cyl}$. For example when the patch is completely wet ($\theta_p=0$) the triple point occurs for
 \bb
a^{\rm cyl}_T=\sqrt{\frac{1-\cos\theta}{1+\cos\theta}}\,.
 \ee
It is interesting to compare this with the result for the slit geometry obtained by setting $\theta_p=0$ in (\ref{triple}) displayed in Fig.~4.
Unlike the slit geometry for which the aspect ratio reaches a maximum value with maximum contrast ($\theta=\pi$) the triple point value of $a_T^{\rm
cyl}$ diverges as $\theta$ is increased to $\pi$. A triple point therefore does not exist for the maximum contrast cylinder. This reflects the fact
that the for the cylindrical geometry the surface tension contributions for the wall-gas and the wall-liquid interfaces scale with the radius
${\mathcal{R}}$. There is no such dependence on the wall separation $L$ for capillary slits.

Finally we turn attention  to possible fluctuation effects which are not captured by macroscopic arguments or our model DFT. Fluctuations do not
effect the direct capillary condensation from a gas-like to a liquid-like state  (occurring for $\theta_E<\theta$ or equivalently $a>a_T$) since the
capillary is pseudo two dimensional. However they do, at least in principle, affect the two other possible transitions in rather different ways. Here
we argue however that they are not quantitative important. Consider the bridging transition first. Since the patch width  $H$ and wall separation $L$
are finite the bridging transition is pseudo one-dimension and standard finite-size scaling arguments determine that this first-order transition is
not precisely sharp but rounded on a scale set by $\exp(-\gamma L H/k_BT)$. This is largest for the bridging transition occurring near the triple
point (since the aspect ratio $a=L/H$ is largest) but even for the smallest systems we have considered here this is negligibly small. For example, at
low temperatures this implies that the transition from a gas-like to bridge state is rounded over the range $\Delta \mu/\mu_{\rm sat}\approx
\exp(-250)$. Fluctuation effects are most subtle for the condensation transition from a bridge phase to a liquid-like state. If the ends of the
capillary are considered open then the transition can be viewed as the unbinding of the menisci from the edges of the patch to the ends of the
capillary. Recall that at the condensation phase boundary, $\theta_E=\theta$, so there is no macroscopic free-energy cost for translating the
meniscus away from the edges of the patch. This allows for the possibility of a second order phase transition. However, more microscopically there
will always be local pinning of the menisci near the patch edges arising from intermolecular forces. If they are strong enough the menisci remain
bound to the patch at $\mu=\mu_{cc}^-$ and the location of the menisci jump discontinuously, via a first-order transition to the ends of the
capillary on the crossing the phase boundary. This is what we observe in the mean-field DFT. In principle, it is possible that thermal fluctuations
in the position of menisci cause it to tunnel away from the patch edge so that its distance from the edge diverges continuously as $\mu$ is increased
to $\mu_{cc}$. This is directly analogous to condensation occurring in a capped capillary above the wetting temperature of the confining walls.
However, since the line tension controlling the fluctuations of menisci (in the $y$ direction)  is large, of order  $\gamma L$ it is unlikely that
such tunnelling occurs and the menisci always remains pinned to the patch edges at $\mu_{cc}$. This transition therefore also always remains
first-order and is accurately described by the classical DFT employed here.

In this paper we have examined the conditions under which bridging precedes capillary condensation in an idealized model of a heterogenous capillary
slit. Using the standard Kelvin and a modified Kelvin equation, which allows correctly for the edge contact of the menisci in the bridge phase, we
have derived a condition for the value of the aspect ratio $L/H$ which  leads to a triple point. Of particular interest is the existence of an upper
universal value for the aspect ratio for a slit with maximum wetting contrast. This mesoscopic approach proves to be extraordinary accurate even for
molecularly narrow slits. Even for walls with single patches a number of extensions of our work are possible. For example, it would interesting to
consider how the stability of the bridge phase is affected when the contact angles of the patches on the two  walls are different and also when the
patches are sheared. This is a topic of future work.

\begin{acknowledgments}
\noindent This work was financially supported by the Czech Science Foundation, Project No. GA17-25100S.
\end{acknowledgments}


\end{document}